\documentclass[twocolumn]{aastex62}

\shorttitle{Number of Satellites vs. Bulge Mass}
\shortauthors{Javanmardi et al.}
\begin{document}

\title{The number of dwarf satellites of disk galaxies versus their bulge mass in the standard model of cosmology}

\correspondingauthor{Behnam Javanmardi}
\email{javanmardi@ipm.ir, behjava@gmail.com}

\author[0000-0002-9317-6114]{B. Javanmardi}
\affil{School of Astronomy, Institute for Research in Fundamental Sciences (IPM), Tehran, 19395-5531, Iran}

\author[0000-0002-1496-3591]{M. Raouf}
\affiliation{School of Astronomy, Institute for Research in Fundamental Sciences (IPM), Tehran, 19395-5531, Iran}

\author[0000-0003-0558-8782]{H. G. Khosroshahi}
\affiliation{School of Astronomy, Institute for Research in Fundamental Sciences (IPM), Tehran, 19395-5531, Iran}

\author[0000-0003-0126-8554]{S. Tavasoli}
%\affiliation{School of Astronomy, Institute for Research in Fundamental Sciences (IPM), Tehran, 19395-5531, Iran}
\affiliation{Faculty of Physics, Kharazmi University, Mofateh Ave., Tehran, Iran}

\author[0000-0003-4552-9808]{O. M{\"u}ller}
\affiliation{Observatoire Astronomique de Strasbourg  (ObAS),
Universite de Strasbourg - CNRS, UMR 7550 Strasbourg, France}

\author[0000-0002-2477-6634]{A. Molaeinezhad}
\affiliation{School of Astronomy, Institute for Research in Fundamental Sciences (IPM), Tehran, 19395-5531, Iran}
\affiliation{Instituto de Astrofísica de Canarias, Calle Vía Láctea s/n, E-38205 La Laguna, Tenerife, Spain}
\affiliation{Departamento de Astrofísica, Universidad de La Laguna, E-38200 La Laguna, Tenerife, Spain}

\begin{abstract}
There is a correlation between bulge mass of the three main galaxies of the Local Group (LG), i.e. M31, Milky Way (MW), and M33, and the number of their dwarf spheroidal galaxies. A similar correlation has also been reported for spiral galaxies with comparable luminosities outside the LG. These correlations do not appear to be expected in standard hierarchical galaxy formation. In this contribution, and for the first time, we present a quantitative investigation of the expectations of the standard model of cosmology for this possible relation using a galaxy catalogue based on the Millennium-II simulation. Our main sample consists of disk galaxies at the centers of halos with a range of virial masses similar to M33, MW, and M31. For this sample, we find an average trend (though with very large scatter) similar to the one observed in the LG; disk galaxies in heavier halos on average host heavier bulges and larger number of satellites. In addition, we study sub-samples of disk galaxies with very similar stellar or halo masses (but spanning a range of 2-3 orders of magnitude in bulge mass) and find no obvious trend in the number of satellites vs. bulge mass. We conclude that while for a wide galaxy mass range a relation arises (which seems to be a manifestation of the satellite number - halo mass correlation), for a narrow one there is no relation between number of satellites and bulge mass in the standard model. Further studies are needed to better understand the expectations of the standard model for this possible relation.

\end{abstract}


\section{Introduction} \label{sec:intro}
In the standard hierarchical structure formation model, galaxies form by condensation and cooling of baryonic matter at the centers of dark matter halos \citep{White1978}. They grow via accretion of more matter and merging with other halos \citep{Mo2010}, and a combination of these two processes are considered to lead to a diverse population of galaxies \citep{Vogelsberger2014}. Every galaxy is considered to be surrounded by numerous smaller dark matter subhalos \citep{Diemand2008}, a considerable fraction of which are expected to host dwarf galaxies. This paradigm has been understood to be successful in explaining the large scale distribution of galaxies \citep[e.g.][]{Eisenstein2005}. However, on small scales and in particular in the Local Group (LG), it is faced with a number of challenges such as the \textit{missing satellites} \citep{Klypin1999}, \textit{too-big-to-fail} \citep{Boylan-Kolchin2011}, and the \textit{disk of satellites} \citep{Kroupa2005, Pawlowski2013} problems. There are ongoing observational, theoretical, and computational efforts to find solutions to these problems \citep[see][for a review]{Bullock2017}. Therefore, studying the intrinsic properties of dwarf satellite galaxies and their relation to their host major galaxy is crucial for further understanding of galaxy formation and fundamental physics \citep[see e.g.][]{Garaldi2018b, Kroupa2018}.

While the dynamical properties of satellite galaxies in the standard model are found to vary for different assembly histories \citep{Garaldi2018}, their overall number mainly depends on the mass of their host halo. The heavier the halo, the larger is the number of its subhalos \citep{Ishiyama2013} and in turn the number of its dwarf satellite galaxies. 

In this standard scenario, bulges of spiral galaxies are generally accepted to form mainly via mergers \citep{Hopkins2010} with larger/smaller bulges tending to appear after major/minor mergers. In addition, disk instabilities can form pseudo-bulges \citep{Bournaud2016}, and if violent, may also form massive bulges \citep{Bell2017}. \citet{Bell2017} reported evidence that mergers may not be the only channel of massive bulge formation, indicating that this could be a more complex process than previously thought \citep[see also][]{Sachdeva2016}.

In a paper on LG tests of dark matter cosmology, \citet[][hereafter K10]{kroupa10} reported a correlation between the bulge mass of the main three galaxies in the LG, i.e. M31, MW, and M33 and the number of their dwarf satellites. They argue that while the number of satellites is expected to be correlated with the rotation velocity (a proxy to the mass of the dark matter halo) of their host spiral galaxies, a correlation between that number and the bulge mass is not expected. The reason is that a correlation between bulge mass and rotation velocity does not appear to exist because galaxies with similar rotation velocity but with and without bulge are observed in nature\footnote{For example M31, M101, NGC3672 and NGC4736 have similar rotation velocities \citep{Faber1979}, but are observed to have very different bulge/total ratios.}. Consequently, galaxies with similar rotation velocities but with and without bulge should have statistically similar number of satellites. Although the correlation in the LG is found for only three galaxies (which would not suffice to accept its ubiquity), and these three galaxies have different dark matter halo masses, it is very important to further study a possible relation between bulge mass and number of satellites. 

In another study, \citet[][hereafter LK16]{Lopez2016} reported a significant ($5\sigma$) correlation between bulge/disk ratio of disk galaxies with similar luminosities (hence similar masses) and number of tidal dwarf satellite galaxies using data from \citet{Kaviraj2012} and \citet{Willett2013} catalogs. LK16 conclude that the correlation they find is consistent with predictions of Milgromian dynamics without dark matter \citep{Milgrom2009} and discuss that such a correlation cannot readily be explained in the standard model. If the majority of the observed satellites are ancient tidal dwarf galaxies (supported by the observed phase-space correlation of dwarf galaxies in MW \citeauthor{Pawlowski2013} \citeyear{Pawlowski2013}, M31 \citeauthor{Ibata2013} \citeyear{Ibata2013}, and Cen A \citeauthor{Mueller2018a} \citeyear{Mueller2018a} groups) then such galaxies without bulges should have statistically far fewer satellites than galaxies with bulges since bulge growth is enhanced significantly in galaxy-galaxy encounters \citep{Kroupa2015}. In addition, \citet{Bilek2018} showed that tidal planes of satellites explaining these phase-space correlations can be formed in Milgromian dynamics.

Both K10 and LK16 emphasize the need for dedicated observational surveys to conclude whether a correlation between bulge mass and number of satellites exists in nature. In fact, investigation of this correlation has been one of the main motivations of the Dwarf Galaxy Survey with Amateur Telescopes \citep[DGSAT,][]{Javanmardi2016} for which \citet{Henkel2017} compiled a catalogue of edge-on disk galaxies with a range of bulge/total ratios. 

Until now, the possible relation between the bulge mass of disk galaxies and the number of their dwarf satellites in the standard model of cosmology has not been investigated in a quantitative way. In this work, we study this relation using the data from the Millennium-II Simulation \citep{Boylan-Kolchin2009} with the semi-analytic galaxy formation model of \citet{Guo2011}. This simulation has enough resolution for studying dwarf satellite galaxies. For example, it has been used by \citet{Bahl2014} and \citet{Ibata2014} to study the disk of satellites around Andromeda-like galaxies in the standard model of cosmology.

This paper is organized as follows. In Section \ref{sec:method} we briefly introduce the Millennium-II simulation and explain our sample selection. In Section \ref{sec:analysis} we present our analysis and  results, and finally in Section \ref{sec:conclusion} we conclude and provide some observational prescriptions for further tests of the main question of this paper.  

\begin{figure*}[th]
    \centering
    \includegraphics[scale=0.29]{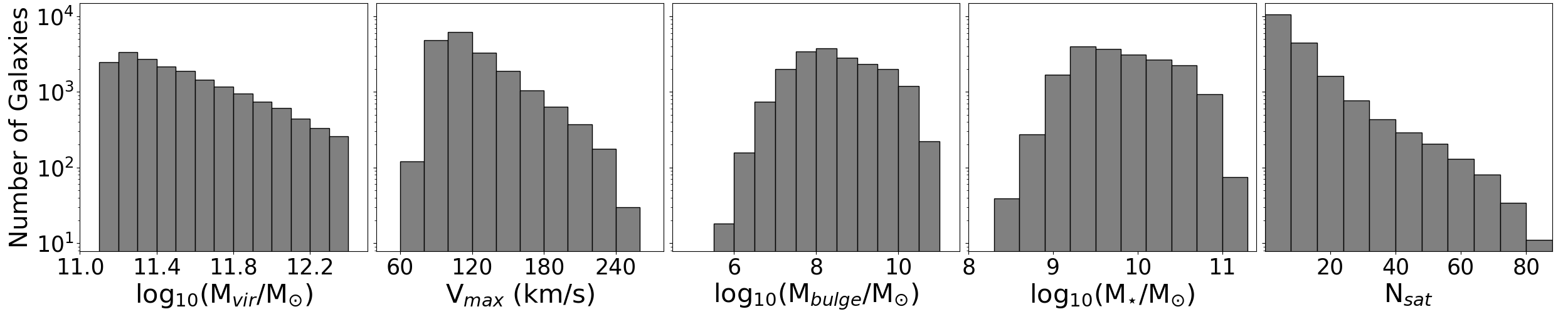}
    \caption{Distribution of virial mass (M$_{vir}$), maximum rotation velocity (V$_{max}$), bulge mass (M$_{bulge}$), and stellar mass (M$_{\star}$) of the disk galaxies in our main sample, and also that of the number of their satellites (N$_{sat}$).}
    \label{fig:hist_4plots_millenium}
\end{figure*}

\section{Method}\label{sec:method}

\subsection{Millennium-II Simulation}
To test the expectations of the standard model for a correlation between bulge mass and number of satellites we use the virtual galaxy catalogue of \citet[][hereafter G11]{Guo2011} which is built on the  Millennium-II cosmological N-body simulation \citep[][hereafter MS-II]{Boylan-Kolchin2009}. It contains 2160$^3$ particles with masses of 9.45$\times10^6$ M$_{\odot}$ and has a box side of 137 Mpc in size. In the MS-II, the values for matter, baryon, and cosmological constant density parameters are $\Omega_M=0.25$, $\Omega_b=0.045$, and $\Omega_{\Lambda}=0.75$, respectively, and $n=1$ for the scalar spectral index, $\sigma_8=0.9$ for the fluctuations amplitude, and $H_0=100 h$ kms$^{-1}$Mpc$^{-1}$ with $h=0.73$ for the Hubble constant.

%The values for the cosmological parameters used by the MS-II are $\Omega_M=0.25$ for the matter density, $\Omega_b=0.045$ for the baryon density, $\Omega_{\Lambda}=0.75$ for the cosmological constant, $n=1$ for the scalar spectral index, $\sigma_8=0.9$ for the fluctuations amplitude, and $H_0=100 h$ kms$^{-1}$Mpc$^{-1}$ with $h=0.73$ for the Hubble constant.

In the MS-II simulation, subhalos down to masses of $2\times10^8$M$_{\odot}$ can be resolved and, as G11 argues, it can be used to study the dwarf satellites of Milky-Way-like galaxies \citep[e.g. Sagittarius dwarf galaxy has a dynimaical mass of $\approx2\times10^8$M$_{\odot}$,][]{McConnachie2012}. Every bound group in the MS-II hosts a central galaxy labeled as type 0 which is located at the most massive subhalo of the group. Type 1 and type 2 galaxies are both satellites of the central galaxy with the difference that type 2 galaxies do not have a resolvable associated dark matter halo \citep[and could be tidal dwarf galaxy candidates, see][for a recent study]{Haslbauer}.

The semi-analytic model of G11 uses three scenarios for bulge formation: minor and major mergers, and disk buckling. In a major merger, all stars end up in a spheroidal component. In a minor merger, the disk survives and the stars of the minor progenitor are added to the bulge. And in secular evolution, when the maximum rotation velocity reaches a certain level (depending on the stellar mass and exponential scale-length of the stellar disc), mass is transferred from the disk to the bulge in order to keep the disk stable. G11 report that their treatment of bulge formation shows good agreement with the observed distribution of galaxy morphological types.

\subsection{Sample Selection}\label{sec:sample}
We choose a sample of galaxies similar in mass to those of M33, MW and M31. The halo mass of M33 is around $10^{11}$M$_{\odot}$ \citep{Seigar2011,Corbelli2000} and that of the M31 is slightly larger than $10^{12}$M$_{\odot}$ \citep{Watkins2010,Tamm2012}. Therefore, we first query all the halos with virial mass $10^{11}\leq\mathrm{M}_{vir} \leq2\times 10^{12}$ M$_{\odot}/h$ from the G11 catalog\footnote{Following G11 we use $h=0.73$ in our analysis.}. Out of these, we select only the halos that have disk galaxies at their centres by checking the ratio of the stellar bulge mass, M$_{bulge}$, to the total stellar mass, M$_{\star}$. Following G11 we use the condition M$_{bulge}$/M$_{\star} < 0.7$ to separate disk galaxies from ellipticals.

In addition, we should exclude from our analysis the halos in which more than one major galaxy resides. For such halos, it is not straightforward to decide on the association of the satellites to each of the major galaxies. To avoid this issue, we reject all the halos in which there exists a non-central (type 1 or 2) galaxy with stellar mass larger than 0.8 times the stellar mass of the central galaxy. This condition also excludes the few halos in which the central galaxy (type 0) does not have the highest stellar mass.

The correlation reported by K10 were obtained by taking into account only the dwarf satellites with V-band luminosity $\mathrm{L}_\mathrm{V}>2\times10^{5}\mathrm{L}_{\odot}$ so it remains robust against later discoveries of fainter satellites. We follow a similar criterion for selection of satellites and by assuming a stellar mass-to-light ratio of 1.5 $\mathrm{M}_{\odot}/\mathrm{L}_{\odot}$ for dwarf galaxies \citep{Dabringhausen2016}, we count only the non-central subhalos with M$_{\star}> 3\times10^{5}\mathrm{M}_{\odot}$. Furthermore, we only count those subhalos that are within the virial radius of the central galaxy. The reason for this is that G11 differentiates between the physical processes affecting the central galaxies and the satellites only when the latter enters the virial radius of the former.

The above conditions provide us with our \textit{main sample} which contains 18646 type 0 halos (or disk galaxies). Figure \ref{fig:hist_4plots_millenium} shows the distribution of virial mass, M$_{vir}$, maximum rotation velocity, V$_{max}$, bulge mass, M$_{bulge}$, and stellar mass, M$_{\star}$, of the disk galaxies in our sample, and also that of the number of their satellites, N$_{sat}$.

\begin{figure*}
    \begin{center}
    \includegraphics[scale=0.35]{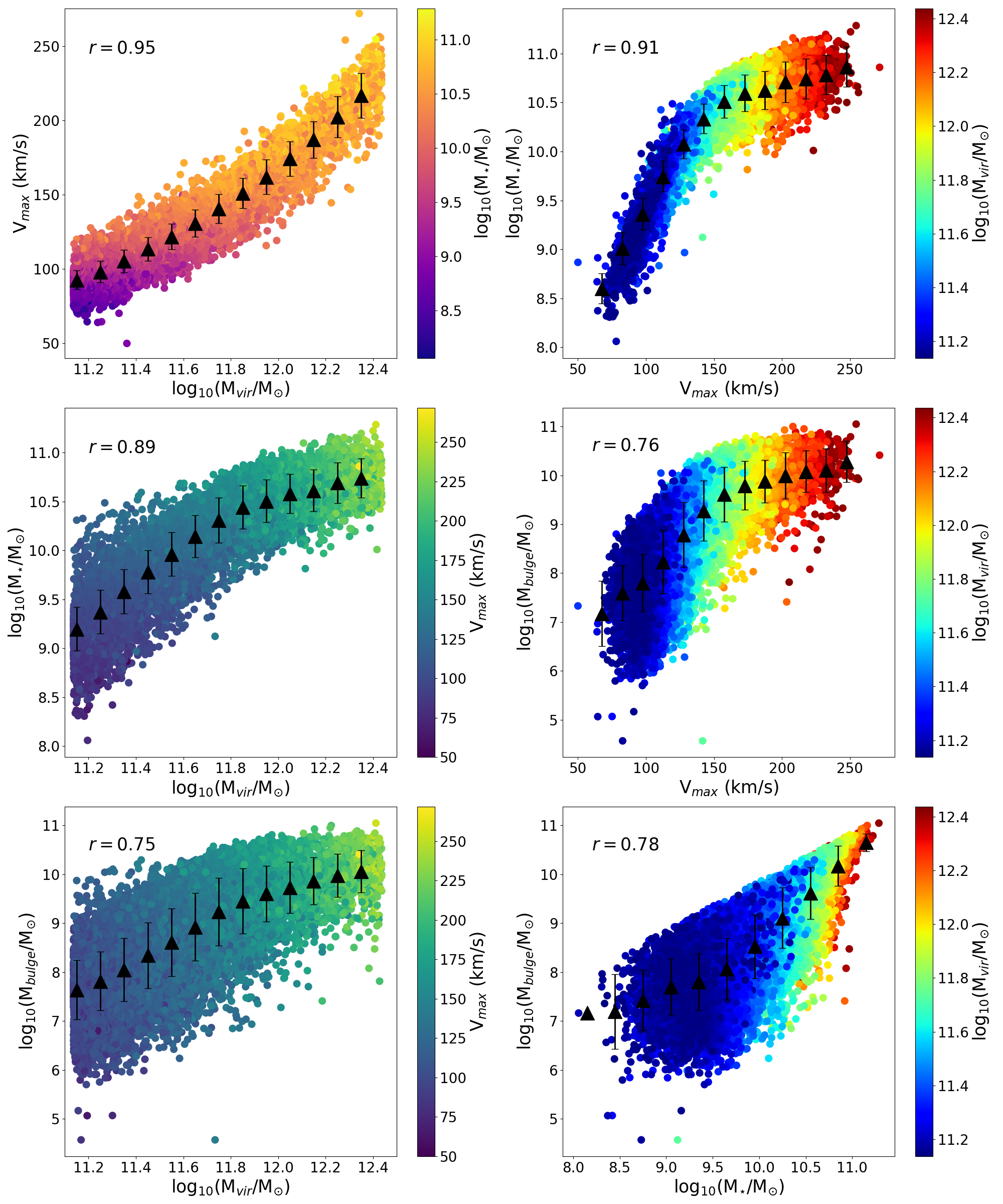}
    \caption{The pairwise relations between M$_{vir}$, V$_{max}$, M$_{\star}$, and M$_{bulge}$ of the galaxies in our main sample, as a check of their expected behaviour. The linear correlation coefficient, $r$, is also shown in each panel. The quantities represented by the color codes are written next to each color bar. Binned mean values and their standard deviations are also shown in each panel to better see the average trends. Note the large scatter in M$_{bulge}$; galaxies with similar values of M$_{vir}$, V$_{max}$, or M$_{\star}$ cover around 3 orders of magnitudes in M$_{bulge}$. The sharp edge at the top of the data distribution in the lower right panel is due to selecting only the disk galaxies with M$_{bulge}$/M$_{\star} < 0.7$ (see Section \ref{sec:sample}).}
    \label{fig:6plots}
    \end{center}
\end{figure*}

\section{Analysis and Results}\label{sec:analysis}
In this section we perform three different analyses. First we study the correlation found in the LG by analyzing the whole main sample, then we analyze sub-samples with similar masses, and in the end we study the number of satellites for \textit{pure-disk} and \textit{bulge-dominated} galaxies.

\subsection{The main sample}\label{sec:main_sample}
\begin{figure*}
    \begin{center}
    \includegraphics[scale=0.35]{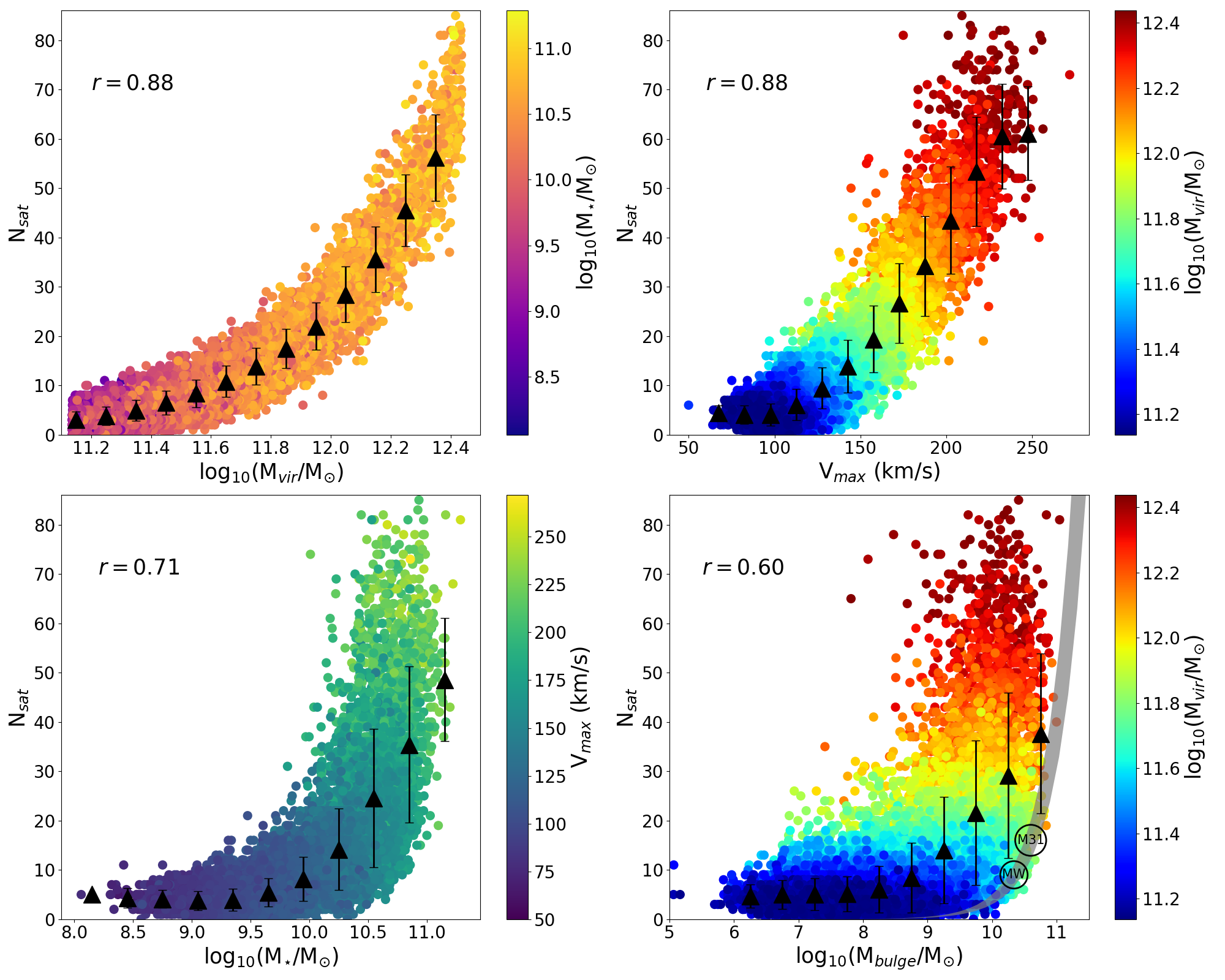}
    \caption{Number of satellites vs. M$_{vir}$ (top left), V$_{max}$ (top right), M$_{\star}$ (bottom left), and M$_{bulge}$ (bottom right). The quantities represented by the color codes are written next to each color bar. Binned mean values and their standard deviations are also shown in each panel to better see the average trends. The linear correlation coefficient $r$ between N$_{sat}$ and each quantity is also shown on each panel. In the N$_{sat}$ vs. M$_{bulge}$ panel we also mark the position of MW and M31, and the grey shaded region is the relation (plus uncertainty) reported by \citet{kroupa10} in their Figure 3. See Section \ref{sec:main_sample} for further explanations.}
    \label{fig:Nsat_plots}
    \end{center}
\end{figure*}

We first study the pairwise relations between M$_{vir}$, V$_{max}$, M$_{\star}$, M$_{bulge}$, and N$_{sat}$ for our main sample. Throughout this study we use the Pearson's linear correlation statistics as an approximation for the level of dependency between each pair of quantities. Some of these relations are very well known and here we check their expected behaviour in our sample. The results are shown in Figures \ref{fig:6plots} and \ref{fig:Nsat_plots}. In each frame of the figures the linear correlation coefficient, $r$, is also shown. This value can be between -1 (fully anti-correlated) and 1 (fully correlated), and zero would indicate no correlation. For all the $r$ values shown in Figures \ref{fig:6plots} and \ref{fig:Nsat_plots}, the p-values (i.e. the probability that such correlation occurs by chance) are found to be vanishingly small (partly due to the large number of data points). At each frame, the binned mean values and their standard deviations are also shown to better see the average trends.

In Figure \ref{fig:6plots}, we see that as expected M$_{\star}$, V$_{max}$, and M$_{vir}$ are strongly correlated with each other; heavier halos host spiral galaxies with larger rotation velocities and larger stellar masses. The important result from this figure is that the bulge mass is also found to statistically scale with M$_{vir}$ (and V$_{max}$). Heavier halos host spirals with more massive bulges, or in other words, bulges form not completely independent of the dark matter halo they sit in. Nevertheless, it should be noted that there is a large scatter (around three orders of magnitude) in M$_{bulge}$ of galaxies with similar V$_{max}$, M$_{vir}$, or M$_{\star}$. 

As can be seen in Figure \ref{fig:Nsat_plots}, N$_{sat}$ shows a clear and strong correlation with M$_{vir}$ and V$_{max}$, heavier halos contain larger number of subhalos. However, for N$_{sat}$ vs. M$_{\star}$ the trend appears not to be monotonic. For small M$_{\star}$ the average number of satellites remains small and does not show a significant change, but from $\log_{10}(\mathrm{M}_{\star}/\mathrm{M}_{\odot})\approx10$, this number shows a rapid increase, though with a large scatter. The flat behaviour of N$_{sat}$ for low stellar masses is most probably related to both the resolution limit and the mass condition we applied to the satellites \citep[see also][]{Puebla2012}. The low M$_{\star}$ halos, that statistically contain a larger number of low M$_{\star}$ subhalos, are most affected by these limits\footnote{We test this by applying two lower mass limits to the satellites, (M$_{\star}/\mathrm{M}_{\odot})>10^{4}$ and $10^{8}$ which move the flattening slightly to the left and to the right, respectively.}. 

The results regarding the main question of this section, is shown in the lower right panel of Figure \ref{fig:Nsat_plots}. We see that N$_{sat}$ vs. M$_{bulge}$ shows a similar behaviour to that vs. M$_{\star}$. For small bulge masses the average number of satellites does not change significantly (which similar to the case for M$_{\star}$ is most probably related to the resolution limit and our satellite mass condition), but from $\log_{10}(\mathrm{M}_{bulge}/\mathrm{M}_{\odot})\approx9$ this number starts to increase with a very large scatter. Nevertheless, a positive correlation is observed and, on average, galaxies with heavier bulges have larger number of dwarf satellites. In the same panel, we also show the correlation found by K10 in the LG. For M33, MW and M31, they used M$_{bulge}/10^{10}\mathrm{M}_{\odot}=0$, 2.2, and 4, respectively. We also adopt the same values as they are in agreement with more recent studies \citep{Seigar2011,Kam2015,Valenti2016, Diaz2017}. We note again that the correlation in K10 takes into account only the satellites with $\mathrm{L}_\mathrm{V}>2\times10^{5}\mathrm{L}_{\odot}$, hence it should not be influenced by the recent discoveries of fainter satellites. Locations of MW and M31 are shown in N$_{sat}$ vs. M$_{bulge}$ panel. M33 would be located on (N$_{sat}$,M$_{bulge}$)=(0,0) which is not shown to prevent the data points from accumulating on one side of the plot. The grey shaded area is the relation (plus its uncertainty) reported by K10 for N$_{sat}$ vs. M$_{bulge}$ in the LG (see their Section 4 and Figure 3). We see that this relation and the mean values in our main sample follow relatively similar trends. The offset in the number might be related to the "missing-satellite-problem". However, studying its effect is beyond the scope of this work as we are mainly interested in the trend rather than the absolute numbers. 

Although the scatter in N$_{sat}$ is large, i.e. different galaxies with similar M$_{bulge}$ have very different numbers of satellites, the overall trend can be understood by remembering that both N$_{sat}$ and M$_{bulge}$ are positively correlated with M$_{vir}$. \citet{Lucia2011} has also report a correlation between the stellar bulge-to-total mass ratio and the halo mass (for halos more massive than $10^{12}$M$_{\odot}$) in the standard model. Our result indicates that a (perhaps indirect) correlation between bulge mass and the number of satellites could emerge in the standard model. Disk galaxies that are located in heavier halos statistically have both heavier bulges and larger number of satellites.\\

\subsection{Galaxies with similar M$_{vir}$, V$_{max}$, or M$_{\star}$}\label{sec:similar_mass}
\begin{figure*}
    \begin{center}
    \includegraphics[width=\textwidth]{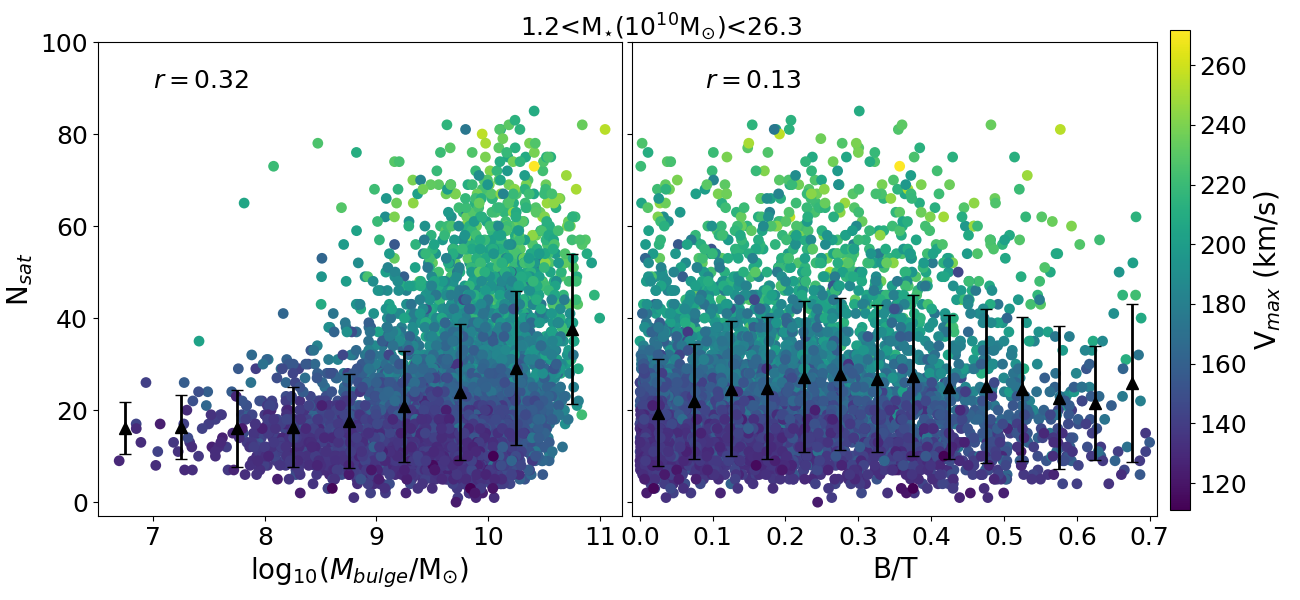}
    \caption{The number of satellites vs. bulge mass (left) and B/T (right) for a sub-sample with the mass range of the \citet{Lopez2016}. Binned mean values and their standard deviations are also shown in each panel to better see the average trends. A weak correlation with bulge mass but no significant correlation with B/T are found. See Section \ref{sec:similar_mass}.}
    \label{fig:lopez-mass}
    \end{center}
\end{figure*}

The significant correlation reported by LK16 was found for spiral galaxies with a mass range narrower than that of our main sample. Galaxies in their sample has baryonic mass in the range $1.2<\mathrm{M}_{\star}/10^{10}\mathrm{M}_{\odot}<26.3$ (or $10.08<\log_{10}(\mathrm{M}_{\star}/\mathrm{M}_{\odot})<11.42$). Applying the exact same condition on the stellar mass of our sample gives us 6042 galaxies. LK16 has used "bulge index" in their analysis. This quantity comes from visual inspection of bulge contribution (ranging from 0 for no bulge to 3 for dominant bulge). A quantity resembling this bulge index in our analysis is bulge to total ratio defined as B/T$\equiv$M$_{bulge}$/M$_{\star}$. For this sub-sample, we measure the correlation of N$_{sat}$ with both bulge mass and B/T. We find that N$_{sat}$ is weakly correlated with $\log_{10}(\mathrm{M}_{bulge}/\mathrm{M}_{\odot})$ with $r=0.32$ (p-value $\approx10^{-116}$) though again with a very large scatter for galaxies with higher masses. However, the correlation coefficient of N$_{sat}$ with B/T is very small $r=0.13$ (p-value $=1.2\times10^{-20}$). This indicates that when the bulge contribution is normalized to the mass of the galaxy, no significant correlation is found. The results are shown in Figure \ref{fig:lopez-mass}. Therefore, it is very important to note that when the galaxies have different masses, a correlation (however negligible) is found between the number of satellites and the mass of the bulge, mainly because galaxies in heavier halos on average grow larger in both total stellar and bulge mass and such halos host a larger number of subhalos.  

\begin{figure*}[t]
    \begin{center}
    \includegraphics[scale=0.3]{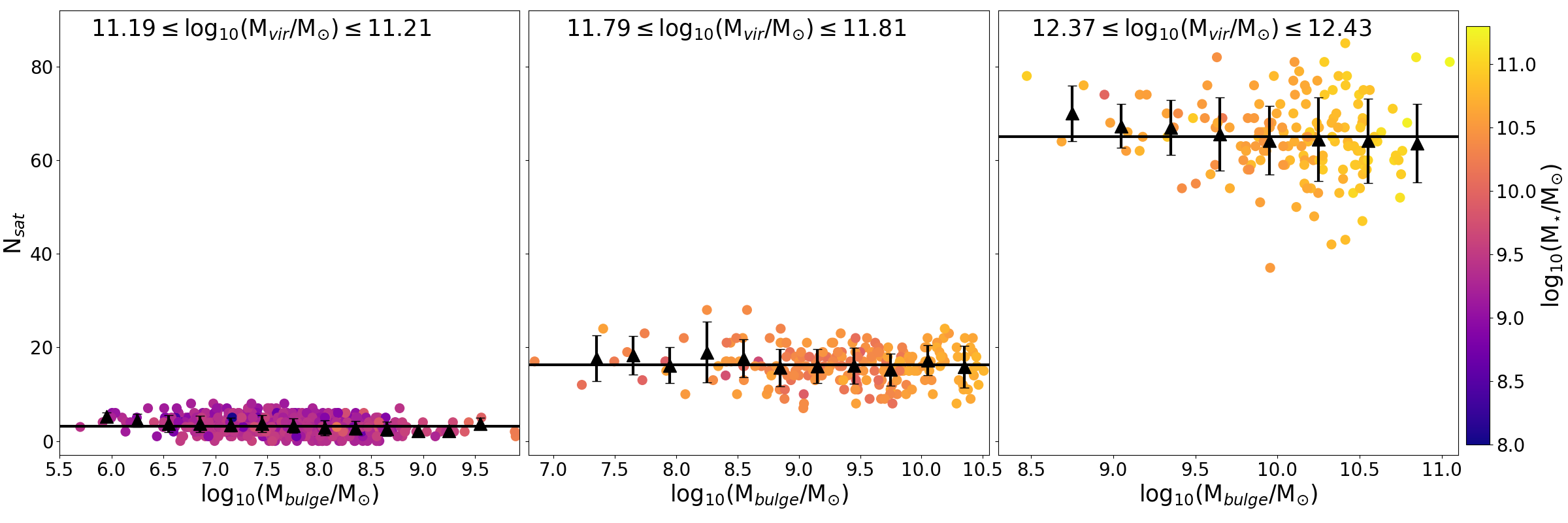}
    \includegraphics[scale=0.3]{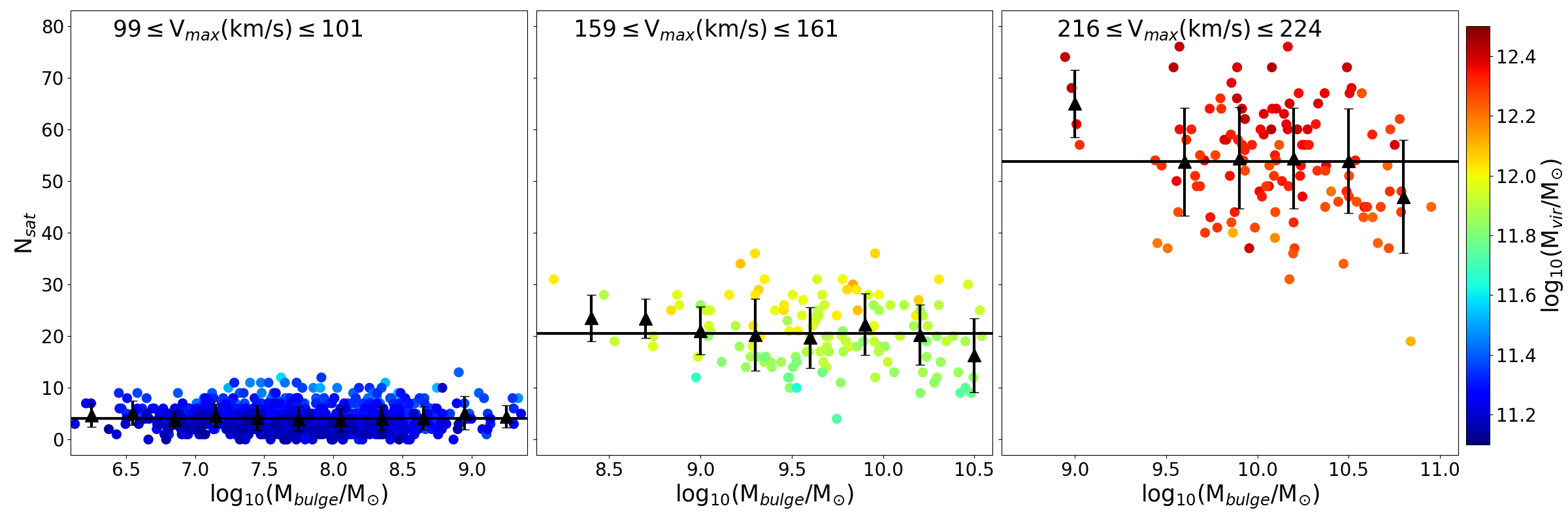}
    \includegraphics[scale=0.3]{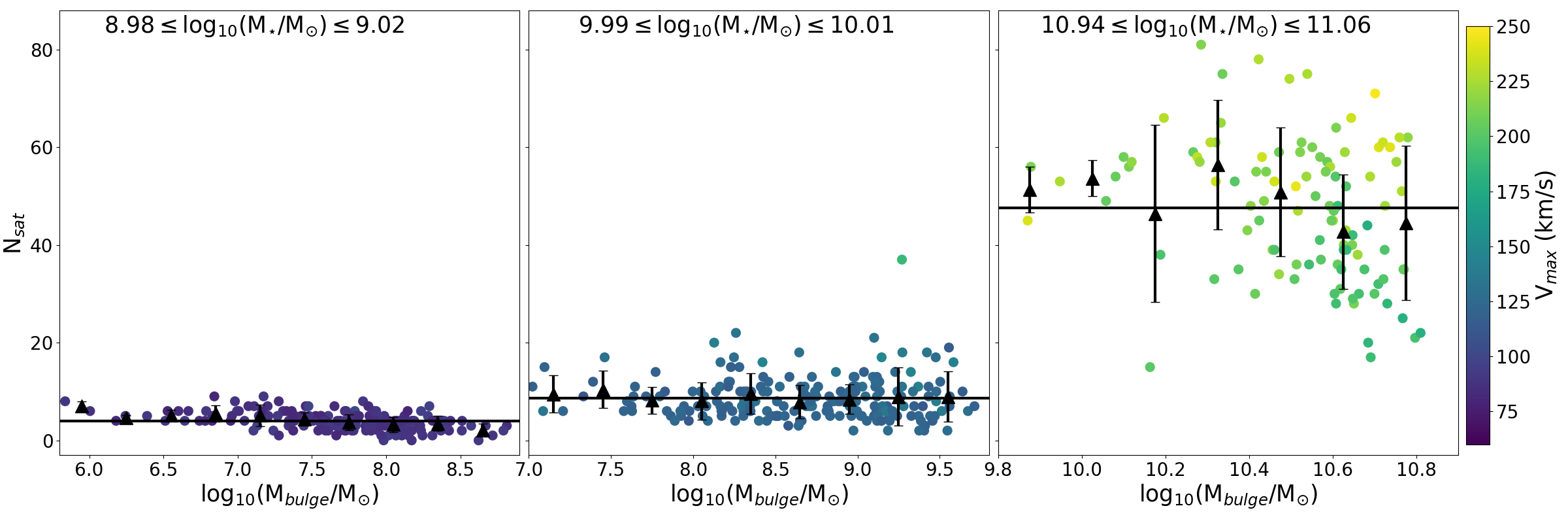}
    \caption{N$_{sat}$ vs. M$_{bulge}$ for sub-samples of galaxies with similar M$_{vir}$, V$_{max}$, and M$_{\star}$. The quantities represented by the color codes are written next to each color bar. The solid black line on each plot represents the mean number of satellites for that sub-sample. binned mean values and their standard deviations are also shown in each panel to better see the average trends. Where the number of galaxies in a sub-sample was small, the range were slightly increased until it contained at least 100 data points. Note that while all sub-samples span a range of 2-3 orders of magnitude in bulge mass, N$_{sat}$ shows a trend-less scatter around the mean values. See also Table \ref{tab:sub_samples}.}
    \label{fig:3plots}
    \end{center}
\end{figure*}

To see the influence of the halo or total stellar mass of galaxies, we probe whether a correlation exists between N$_{sat}$ vs. M$_{bulge}$ for galaxies with very similar M$_{vir}$, V$_{max}$, or M$_{\star}$. The narrow ranges for each of these quantities are chosen in three distinct regimes, low, intermediate, and high masses or velocities. Figure \ref{fig:3plots} and Table \ref{tab:sub_samples} present the results for nine narrow ranges of M$_{vir}$, V$_{max}$, and M$_{\star}$. Where the number of galaxies in a sub-sample was small, the range were slightly increased until it contained at least 100 data points. On each panel, the black solid line represents the mean number of satellites ($\bar{N}_{sat}$) for that sub-sample, also listed in Table \ref{tab:sub_samples}. Similar to previous plots, the binned mean values are also shown to see the trends (if any). In Table \ref{tab:sub_samples}, we also list the number of disk galaxies, N$_{gal}$, in each sub-sample, the linear correlation coefficient, $r$, between N$_{sat}$ and M$_{bulge}$, and the p-value for each $r$ (i.e. the probability that such anti/correlation occurs by chance). 
\begin{table}[t]
    \centering
    \begin{tabular}{lcccc}
    %\hline
    sub-sample&N$_{gal}$&$r$&p-value& $\bar{N}_{sat}$\\
    \hline
    $\log_{10}(\mathrm{M}_{vir}/\mathrm{M}_{\odot})$ &\\
    $[11.19,11.21]$& 727 & -0.25 & $5.5\times10^{-12}$ & 3.2\\
    $[11.79,11.81]$& 200 & -0.08 & 0.25 & 16.2\\
    $[12.37,12.43]$& 154 & -0.13 & 0.11 & 65.0 \\
    \hline
    $\mathrm{V}_{max}$(km/s) & \\
    $[99,101]$ & 823 & -0.02 & 0.51 & 4.1\\
    $[159,161]$ & 134 & -0.16 & 0.06 & 20.5\\
    $[216,224]$ & 122 & -0.28 & 0.002 & 53.8\\
    \hline
    $\log_{10}(\mathrm{M}_{\star}/\mathrm{M}_{\odot})$ &\\
    $[8.98,9.02]$& 169 & -0.45 & $5.3\times10^{-10}$ & 3.9\\
    $[9.99,10.01]$& 203 & -0.004 & 0.95 & 8.7\\
    $[10.94,11.06]$& 106 & -0.34 & 0.0003 & 47.5\\
    \hline
    \end{tabular}
    \caption{Sub-samples with narrow ranges of M$_{vir}$, V$_{max}$, and M$_{\star}$. N$_{gal}$ is the number of central galaxies, $r$ is the linear correlation coefficient between N$_{sat}$ and M$_{bulge}$, p-value is the probability that such anti/correlation occurs by chance, and $\bar{N}_{sat}$ is the mean number of satellites for each sub-sample. See also Figure \ref{fig:3plots}.}
    \label{tab:sub_samples}
\end{table}

Most of the $r$ values are very small and also no obvious trend is seen for any of the sub-samples. For low (halo and stellar) mass galaxies, small but significant (with small p-value) anti-correlations are found. This could be understood by noting that for these regimes the rate of increase in number of satellites is smaller than the growth rate of bulge mass (see Figures \ref{fig:6plots} and \ref{fig:Nsat_plots}). For the rest of sub-samples, no significant correlation is found. As expected from the analysis in Section \ref{sec:main_sample}, the the average number of satellites, and also its scatter, increases with increasing M$_{vir}$, V$_{max}$, and M$_{\star}$. The scatter is largest for high stellar mass sub-sample. The main conclusion from this analysis is that, despite M$_{bulge}$ spans a range of 2-3 orders of magnitudes in these sub-samples, no correlation is found between N$_{sat}$ vs. M$_{bulge}$ at fixed M$_{vir}$, V$_{max}$, or M$_{\star}$.

\begin{figure*}[t]
    \begin{center}
    \includegraphics[width=\textwidth]{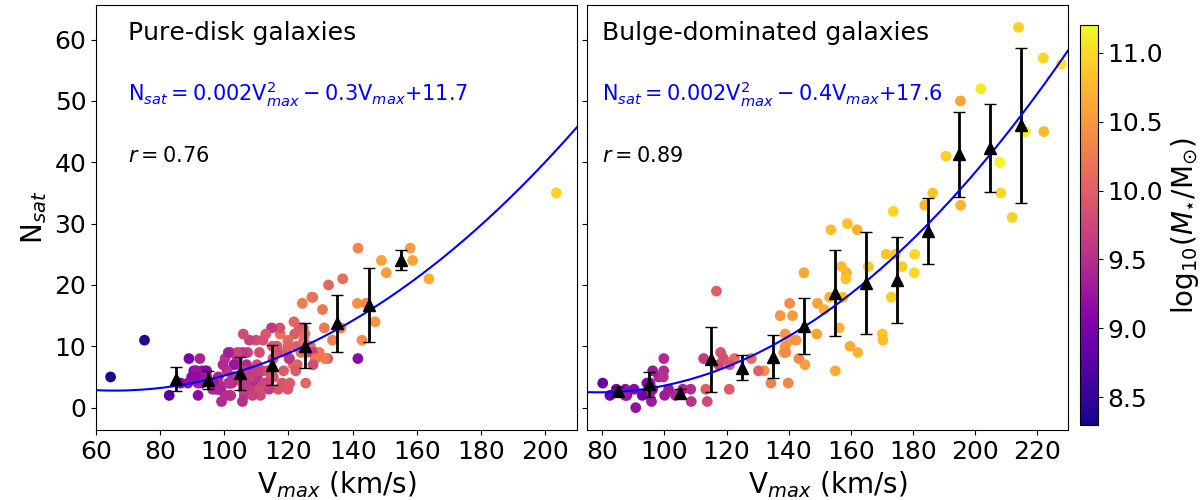}
    \caption{N$_{sat}$ vs. V$_{max}$ with color code being M$_{\star}$ for pure-disk (left) and bulge-dominated (right) galaxies. Binned mean values and their standard deviations are also shown in each panel to better see the average trends. The number of satellites increases with V$_{max}$ and M$_{\star}$. The solid curve is a second degree polynomial fit to the data. The fitting results and the correlation coefficients are also written on each panel. The behaviour of N$_{sat}$ vs. V$_{max}$ is very similar for both types of galaxies. See Section \ref{sec:pure-disk}.}
    \label{fig:pure-disk}
    \end{center}
\end{figure*}
\subsection{Pure-disk and bulge-dominated galaxies}\label{sec:pure-disk}
If a physical relation (i.e. a causal and not an indirect correlation) exists between the number of satellites of a spiral galaxy and the mass of its bulge, all galaxies with negligible bulges (or \textit{pure-disk} galaxies) are expected to have very few or zero number of satellites \citep[similar to M33,][]{kroupa10}. In this section, we consider two sub-samples of galaxies, one with very small and another with large bulge/total ratios. G11 refers to galaxies with M$_{bulge}$/M$_{\star} < 0.03$ as \textit{pure-disk} galaxies. Here, we apply a very conservative cut and select (from our main sample) galaxies with M$_{bulge}$/M$_{\star} < 0.001$ as \textit{pure-disks}. Around 1\% of our main sample (161 galaxies) satisfy this condition. For \textit{bulge-dominated} sample we select galaxies with M$_{bulge}$/M$_{\star} > 0.6$ which constitutes 92 galaxies of our main sample. The \textit{pure-disk} galaxies have halo masses in the range $11.1\lesssim\log_{10}(\mathrm{M}_{vir}/\mathrm{M}_{\odot})\lesssim12.2$, stellar mass in the range $8.4\lesssim\log_{10}(\mathrm{M}_{\star}/\mathrm{M}_{\odot})\lesssim10.9$, and maximum rotation velocity in the range $65\lesssim$V$_{max}$(km/s)$\lesssim204$. These ranges for the \textit{bulge-dominated} galaxies are $11.1\lesssim\log_{10}(\mathrm{M}_{vir}/\mathrm{M}_{\odot})\lesssim12.4$, $8.8\lesssim\log_{10}(\mathrm{M}_{\star}/\mathrm{M}_{\odot})\lesssim11.2$, and $80\lesssim$V$_{max}$(km/s)$\lesssim228$. In Figure \ref{fig:pure-disk} we show N$_{sat}$ vs. V$_{max}$ for these two sub-samples with color code being M$_{\star}$. It is seen that N$_{sat}$ increases with V$_{max}$ for both types of galaxies. We fit second degree polynomials to both samples and the fitting results are also written on each panel. We see that for both samples the behaviour of N$_{sat}$ vs. V$_{max}$ is very similar and at fixed V$_{max}$, galaxies of both samples have similar N$_{sat}$. Only when averaged over all V$_{max}$, the \textit{bulge-dominated} galaxies have a larger N$_{sat}$ (because they are mostly found in heavier halos, see Section \ref{sec:main_sample}). The linear correlation coefficient, $r$, is also written on both panels (the p-values for both of $r$ values are around $10^{-32}$). We conclude that in the standard model, regardless of the bulge mass, the number of satellites increases in a similar way with V$_{max}$ (a proxy to the halo mass) for both types of galaxies. In particular for \textit{pure-disk} galaxies, N$_{sat}$ ranges from 1 to 35 with a mean value of $\approx8$ and the most probable value of $4$. This indicates that in the standard hierarchical structure formation model, as intuitively expected \citep{Kroupa2012}, even \textit{pure-disk} galaxies do have satellites and their average number increases with the mass of the galaxy.

\section{Summary and conclusion}\label{sec:conclusion}
The observed correlation between bulge mass of M33, MW, and M31 in the LG and the number of their satellites, and a similar correlation outside the LG were proposed by \citet{kroupa10} and \citet{Lopez2016} to be a possible challenge for the standard paradigm of galaxy formation and in agreement with Milgromian dynamics without dark matter \citep{Milgrom2009}. In this work, and for the first time, we investigated the expectations from the standard model of cosmology for this possible relation. We used the semi-analytic galaxy formation model of \citet{Guo2011} applied to the Millennium-II Simulation \citep{Boylan-Kolchin2009}. From this catalog, we selected halos with mass in the range $10^{11}\leq\mathrm{M}_{vir} \leq2\times 10^{12}$ M$_{\odot}/h$ (with $h=0.73$) that have disk galaxies at their centre with M$_{bulge}$/M$_{\star} < 0.7$, and that do not have a non-central galaxy with M$_{\star}>0.8\times$M$_{\star,central}$. For satellite galaxies, we considered only the subhalos that have M$_{\star}> 3\times10^{5}\mathrm{M}_{\odot}$, and that are located inside the virial radius of the central galaxy. We investigated the relation between N$_{sat}$ and M$_{bulge}$ for i) the above sample, and ii) for nine sub-samples with narrow ranges of M$_{vir}$, V$_{max}$, or M$_{\star}$. In addition, we studied the N$_{sat}$ for \textit{pure-disk} and \textit{bulge-dominated} galaxies. The main results are as follows:

\begin{enumerate}
    \item We find that although there is a large scatter in the N$_{sat}$ for any given M$_{bulge}$, a statistical trend emerges and disk galaxies with heavier bulges have, on average, a larger number of satellites. The average trend we find for this sample of virtual galaxies is similar to the relation found by \citet{kroupa10} for the LG (except for the case of M33 with negligible bulge and no confirmed satellites). Based on this, we conclude that for a sample of galaxies with different halo mass, those in heavier halos have, on average, both heavier bulges and a larger number of satellites.
    
    \item For sub-samples of galaxies with similar M$_{vir}$, V$_{max}$, or M$_{\star}$ the situation is completely different. All these sub-samples span 2-3 orders of magnitudes in M$_{bulge}$ and in none of them an obvious trend is seen in N$_{sat}$ vs. M$_{bulge}$. From this finding we conclude that galaxies with similar masses can host bulges with very different masses and the number of their satellites appears to be completely independent of that. We also specifically investigated a sub-sample with the mass range of that of \citet{Lopez2016} in which we found no significant correlation between N$_{sat}$ and B/T. This indicates that the $5\sigma$ correlation reported in that study, if confirmed by further observations, could be a challenge for the standard hierarchical structure formation. However, we note that the bulge indices used by \citet{Lopez2016} were assigned by visual inspection which may introduce observational biases.
    
    \item For two sub-samples of \textit{pure-disk} and \textit{bulge-dominated} galaxies (spanning a wide range in M$_{vir}$, V$_{max}$, and M$_{\star}$), we find a strong and similar correlation between N$_{sat}$ and V$_{max}$ (a proxy to halo mass). In particular, we found that \textit{pure-disk} galaxies can be formed in halos with very different masses and can have very different numbers of satellites. In fact, in the standard model, all of them are expected to have at least a few satellites.
\end{enumerate}

Additional studies using other semi-analytic models \citep[e.g.][]{Raouf2017} and hydrodynamic cosmological simulations \citep[such as the Illustris,][]{Vogelsberger2014} are needed to better understand the expectations from the standard model. In addition, further investigations of the observational properties of bulges \citep[e.g.][]{Khosroshahi2000,Molaeinezhad2016,Molaeinezhad2017} would be needed for a better comparison of different models. Based on our results we can prescribe the following observational studies for detailed investigations of the N$_{sat}$ vs. M$_{bulge}$ relation.

\begin{itemize}
    \item To test the correlation found in the LG \citep{kroupa10}, the number of known satellite galaxies around disk galaxies outside the LG should be increased. A systematic survey of dwarf galaxies around disk galaxies with different total and bulge mass would enable testing the correlation observed in the LG.
    
    \item To test the correlation found for galaxies with similar masses or luminosities \citep{Lopez2016}, two types of carefully selected disk galaxy samples can be used; one with similar V$_{max}$ and the other with similar M$_{\star}$. These galaxies must obviously have a wide range of M$_{bulge}$. In fact, for the DGSAT project, \citet{Henkel2017} have already compiled a catalog of more than 30 edge-on disk galaxies with similar luminosities and with distances $\lesssim40$ Mpc for the exact same purpose.
    
    \item An additional observational study could be to select a sample of \textit{pure-disk} galaxies with a wide range of V$_{max}$, and survey their surroundings for dwarf satellites. If there is a physical relation between N$_{sat}$ and M$_{bulge}$, the majority of these galaxies should have zero or only a few satellite galaxies. On the other hand, if such an study results in a correlation between N$_{sat}$ and V$_{max}$ (or M$_{\star}$), that would be consistent with the standard hierarchical structure formation.
\end{itemize}
As a matter of fact, there already have been many surveys of dwarf galaxies outside the LG \citep[e.g.][]{Javanmardi2016,Merritt2016,Ordenes2016,Mueller2017,Mueller2017b,Geha2017,Bennet2017,Danieli2018,Mueller2018,Greco2018,Prole2018,Xi2018,Taylor2018,Martinez-Delgado2018}. Such surveys would eventually enable more detailed studies of the small scale challenges of the standard galaxy formation paradigm, and in particular the relation studied in this work. 

\acknowledgments
We would like to thank Pavel Kroupa, Cristiano Porciani, and also the anonymous referee for their constructive comments. 

\bibliography{biblio}

\end{document}